\documentclass[a4paper,12pt]{article}
\usepackage[left=2.5cm,bottom=3cm,right=2.5cm,top=3cm]{geometry} 

\usepackage{overpic,youngtab}
\usepackage{subfigure}
\usepackage[latin,english]{babel}
\usepackage{amsmath}
\usepackage{amssymb}
\usepackage{epstopdf}
\usepackage{graphics,psfrag,rotating}
\usepackage{mathabx}
\usepackage{graphicx}
\usepackage{dcolumn}
\usepackage{float}
\usepackage{pdflscape}
\usepackage{array}
\usepackage{booktabs}
\usepackage{amscd} 
\usepackage{mathtools}
\usepackage{fancybox}
\usepackage{fix-cm}
\usepackage[colorlinks=true,citecolor=blue,,linktocpage=true,linkcolor=blue,urlcolor=black]{hyperref}
\usepackage{tikz} 
\usetikzlibrary{matrix} 
\usetikzlibrary{positioning} 
\usetikzlibrary{arrows}
\usepackage{titlesec}
\usepackage{abstract}

\def\be{\begin{equation}}
\def\ee{\end{equation}}
\def\bea{\begin{eqnarray}}
\def\eea{\end{eqnarray}}
\def\pd{\partial}
\def\a{\alpha}
\def\b{\beta}
\def\g{\gamma}
\def\d{\delta}
\def\m{\mu}
\def\n{\nu}

\def\l{\lambda}

\def\r{\rho}

\def\s{\sigma}
\def\e{\epsilon}
\def\bi{\begin{itemize}}
\def\ei{\end{itemize}}

\newcommand{\email}[1]{\href{mailto:#1}{\tt #1}}

\begin{document}

		\vspace*{-1cm}
		\phantom{hep-ph/***} 
		{\flushleft
			{{FTUAM-16-37}}
			\hfill{{ IFT-UAM/CSIC-16-102}}}
		\vskip 1.5cm
		\begin{center}
		«{\LARGE\bfseries Weyl Gravity Revisited}\\[3mm]
			\vskip .3cm
		
		\end{center}
		\vskip 0.5  cm
		\begin{center}
			{\large Enrique \'Alvarez}~$^{a}$
			{and \large Sergio Gonz\'alez-Mart\'in}~$^{a}$,
			\\
			\vskip .7cm
			{\footnotesize
				$^{a}~$Departamento de F\'isica Te\'orica and Instituto de F\'{\i}sica Te\'orica, 
				IFT-UAM/CSIC,\\
				Universidad Aut\'onoma de Madrid, Cantoblanco, 28049, Madrid, Spain\\
				\vskip .1cm

				\vskip .5cm
				\begin{minipage}[l]{.9\textwidth}
					\begin{center} 
						\textit{E-mail:} 
						\email{enrique.alvarez@uam.es},
						\email{sergio.gonzalez.martin@csic.es},
					\end{center}
				\end{minipage}
			}
		\end{center}
	\thispagestyle{empty}
	
\begin{abstract}\vspace{-1em}
	\noindent
The on shell equivalence of first order and second order formalisms for the Einstein-Hilbert action does not hold for those actions quadratic in curvature. It would seem that by considering the connection and the metric as independent dynamical variables, there are no quartic propagators for any dynamical variable. This suggests that it is possible to get both renormalizability and unitarity along these lines. We have studied a particular instance of those theories, namely Weyl gravity. In this first paper we show that it is not possible to implement this program with the Weyl connection alone.
\end{abstract}

\newpage
\tableofcontents
	\thispagestyle{empty}
\flushbottom

\newpage
\section{Introduction.}

It is well-known that general relativity is not perturbatively renormalizable (cf.\cite{Alvarez} and references therein for a general review). However, quadratic (in curvature) theories are renormalizable, albeit not unitary \cite{Stelle}, when considered in second order formalism.
\par
Nevertheless, it is possible to let the affine connection be independent from the riemannian metric in a manifold. In the usual first order Palatini \cite{Palatini} approach (linear in curvature), the connection and the metric are treated as independent variables and the Levi-Civita connection appears only when the equations of motion are used. Theses equations of motion enforce that the covariant derivative of the metric should vanish, which means that the connection must be the Levi-Civita one in the torsionless case. Quantization of this Palatini action through the background field method \cite{Buchbinder} is claimed to be essentially equivalent to the Einstein-Hilbert one, first performed in a classic paper by 't Hooft and Veltman \cite{tHooft}.
\par
When more general quadratic in curvature metric-affine actions are considered \cite{Sotiriou}\cite{Wheeler}\cite{Korzynski} the deterministic relationship between the affine connection and the Levi-Civita one is lost, even on shell. That is, the equations of motion do not force the connection to be the Levi-Civita one. This is interesting because it appears that we could  have all the goods of  quadratic lagrangians \cite{Stelle} (mainly renormalizability) without conflicting with the spectral theorem of K\"allen-Lehmann, which  guarantees that in a unitary theory, and in flat space
\be
\langle \Omega|T \phi(x)\phi(y)|\Omega\rangle=i\int~{d^4p\over (2\pi)^4}~e^{ip(x-y)}~\int_0^\infty d \m^2 {\r (\m^2)\over p^2-\m^2+i\e}
\ee
where the {\em spectral function} $\r(\m^2)\geq 0$ is positive semidefinite.
\par
 It follows that it is not possible in a unitary quantum field theory to have propagators falling off at infinity faster than ${1\over k^2}$ with a positive spectral function $\r(m^2)$. On the down side, if the connection is to be a really independent physical variable, we should be able to find its physical meaning insofar as it is not fully determined by the spacetime metric itself. 
\par
What has been said in the preceding paragraph  is a simple consequence of the fact that given any action (suppressing indices)
\be
S\left[\Gamma,g\right]
\ee
in which the dependence on the connection has been separated from the dependence on the metric, the total variation with respect to the metric can be written as
\be
\d_2 S=\int d(vol_x)\bigg\{d(vol)_y {\d S\over \d \Gamma(y)}{\d \Gamma(y)\over \d g(x)}+{\d S\over \d g(x)}\bigg\}\d g(x)
\ee
Therefore, the second order variations are the total functional differential, whereas the first order variation is the set of partial functional derivatives.
\par
It is the clear that the vanishing of the first order variations
\bea
&&{\d S\over \d g_{\a\b}}=0\nonumber\\
&&{\d S\over \d \Gamma^\a_{\b\g}}=0
\eea
implies the vanishing of the second order ones
\be
\d_2 S=0
\ee
but no the other way round; it is a distinct logical possibility to have $\d_2 S=0$ through a cancellation of two non-vanishing terms in the above equation.
\par
The purpose of the present paper is to explore some of the possibilities opened up by this framework.
\par
In this sense, we shall consider connections on the {\em frame bundle} \cite{Kobayashi}\cite{Sharpe}, namely the principal bundle associated to the tangent bundle, with structure group $SO(3,1)$ or $SO(n)\subset GL(n,\mathbb{R})$, in the euclidean case). For simplicity, we shall restrict ourselves to torsionless connections (the observationally allowed parameter space for the torsion is anyway quite thin \cite{Shapiro}). We will write all formulas in the minkowskian signature although as usual all determinants are defined in a riemannian setting and analytical continuation is to be performed afterwards. 
\par
An orthonormalized frame will be characterized by n differential forms
\be
 e^a\equiv e^a_\m dx^\m
\ee
$a=1\ldots n$ are tangent (Lorentz) indices, and $\m,\n\ldots$ are spacetime (Einstein) indices. They obey
\be
\eta_{ab}~e^a_\m(x)~ e^b_\n(x) = g_{\m\n}(x)
\ee
Spacetime tensors are observed in the frame as spacetime scalars, id est,
\be
V^a(x)\equiv e^a_\m(x)~ V^\m(x)
\ee
\par
The Lorentz (usually called spin) connection is defined by demanding local Lorentz invariance of  derivatives of such scalars as
\be
\nabla_\m V^b\equiv \pd_\m V^b+\omega_\m\,^b\,_c V^c
\ee

Physical consistency demands that the Lorentz and Einstein connections are equivalent, that is, that
\be
\nabla_a V^b= e_a^\m e_{b\r} \nabla_\m V^\r
\ee
This is easily seen to imply that
\be
\omega_{abc}=-e_c^\r\pd_a e_{b\r}+\eta_{bd}\Gamma^d_{ac}
\ee
showing that Lorentz and Einstein connections are equivalent assuming knowledge of the frame field (tetrad).
\par
The Riemann Christoffel tensor is completely analogous to the usual gauge non-abelian field strength. The main difference between the curvature tensor and the non-abelian field strength stems from the torsionless algebraic Bianchi identity
\be
R^a\,_b\wedge e^b=0
\ee
which is the origin of the symmetry between Lorentz and Einstein indices
\be
R_{\a\b\g\d}\equiv e_\a^a e_\b ^b~R_{a b \g\d}= R_{\g\d\a\b}\equiv e_\g^c e_\d^d R_{c d \a\b}
\ee
This identity  does not have any analogue in a non abelian gauge theory in which these two sets of indices remain unrelated. The opposite happens with the differential Bianchi identity
\be
d R^a\,_b+R^a\,_c\wedge \omega^c\,_b-\omega^a\,_c\wedge \Omega^c\,_b=0
\ee
which still holds for non-abelian gauge theories when the gauge group is not identified with the tangent group. We have included in the appendix a general treatment of (non Levi-Civita) torsionless Lorentz connections.\newline

The connections is assumed to be inert under Weyl rescalings
\be
g_{\m\n}\rightarrow \Omega^2 g_{\m\n}
\ee\par
 this implies that the Riemann-Christoffel tensor is Weyl invariant as well
\be
R^\m_{~\n\r\s}\rightarrow R^\m_{~\n\r\s}
\ee\par
and
\be
R_{\m\n\r\s}R^{\m\n\r\s}\rightarrow \Omega^{-4}~R_{\m\n\r\s}~R^{\m\n\r\s}
\ee\par
With this definition, the action
\be
S=\int \sqrt{-g}~R_{\m\n\r\s}R^{\m\n\r\s}d^4 x\label{riemannaction}
\ee
transforms as $\Omega^{n-4}$; so that is Weyl invariant in four dimensions.\par
 Since we are considering the Riemann-Christoffel tensor  as a function of an arbitrary connection,
\be
R[\Gamma]^\m_{~\n\r\s}\equiv \pd_\r\Gamma^\m_{\n\s}-\pd_\s\Gamma^\m_{\n\r}+ \Gamma^\m_{\l\r}\Gamma^\l_{\n\s}-\Gamma^\m_{\l\s}\Gamma^\l_{\n\r} 
\ee  
in addition of transforming as a true tensor under arbitrary diffeomorphisms, it may seem that it has an abelian gauge invariance under
\be
\Gamma^\a_{\b\g}\rightarrow \Gamma^\a_{\b\g}+\d^\a_\b~\pd_\g \Omega(x)
\ee
This invariance is not a true symmetry however. Actually what happens is that the transformed connection is not symmetric, which in turn  means that the transformed field generates torsion. Indeed, under a symmetric transformation
\be
\Gamma^\a_{\b\g}\rightarrow \Gamma^\a_{\b\g}+\d^\a_\b~\pd_\g \Omega(x)+\d^\a_\g~\pd_\b \Omega(x)
\ee
the Riemann tensor transforms as
\bea
&&R[\Gamma]^\m_{~\n\r\s}\rightarrow R[\Gamma]^\m_{~\n\r\s}+\d^\m_\r~\left(\nabla_\s\nabla_\n~\Omega-\nabla_\n~\Omega~\nabla_\s~\Omega\right)-\nonumber\\
&&-\d^\m_\s\left(\nabla_\n \nabla_\r ~\Omega-\nabla_\r \Omega~\nabla_\n \Omega\right)
\eea
The extra piece  does not vanish in general. Again, the reason for this is that now the first pair of indices of the Riemann tensor are not related to the second pair on indices as is the case for the Levi-Civita connection owing to the algebraic Bianchi identity.
\par
We have just pointed out that  the nonmetricity is non-vanishing the Riemann tensor does not enjoy the usual symmetries
\be
R[\Gamma]_{\m\n\r\s}\neq R[\Gamma]_{\r\s\m\n}
\ee
as well as
\be
R[\Gamma]_{(\m\n)\r\s}\neq 0
\ee
There are then two different traces. The one that corresponds to the Ricci tensor
\be
R^+[\Gamma]_{\n\s}= g^{\m\r}~R[\Gamma]_{\m\n\r\s}
\ee
and a different one
\be
R^-[\Gamma]_{\m\s}\equiv g^{\n\r}R[\Gamma]_{\m\n\r\s}
\ee
Neither of them is in general symmetric now. There is also an antisymmetric further trace
\be\label{ricciblando}
{\cal R}_{\r\s}\equiv g^{\m\n}~R[\Gamma]_{\m\n\r\s}
\ee
It is a fact that
\be
R^+\equiv g^{\m\n} R^+_{\m\n}=-R^-\equiv g^{\m\n} R^-_{\m\n}\label{traces}
\ee

\section{General non-Levi-Civita torsionless connection}

 Define the {\em non-metricity tensor} (NM) as the covariant derivative of the metric tensor
\be
\nabla_c \eta_{a b}\equiv -Q_{c ab}=-\omega_c\,^d\,_a\eta_{db}-\omega_c\,^d\,_b\eta_{ad}=-\omega_{cba}-\omega_{cab}
\ee	
The symmetric piece of the connection is then precisely
\be
\omega_{a(bc)}=2 Q_{abc}
\ee
The frame field is essentially characterized by its structure constants of the frame. Those  are defined as usual as 
\be
\left[e_a,e_b\right]=C_{ab}^c e_c
\ee
and 
\be
C_{c a b}\equiv \eta_{ce} C^e_{ab}=\g_{bca}-\g_{acb}
\ee

Indeed, the vanishing of the torsion tensor 
\be
d e^a+\omega^\a\,_b\wedge e^c=0
\ee
yields the missing antisymmetric piece of the Lorentz connection $\omega_{a[bc]}$
(remember that the symmetric piece was determined by the non-metricity)
\be
\omega_{abc}-\omega_{acb}=C_{abc}
\ee
The general torsionless connection is then determined in terms of the non-metricity and the structure constants of the frame field as
\be
\omega_{abc}=Q_{abc}+{1\over 2} C_{abc}
\ee

\newpage
\section{Metric from connection}
Given any connection, it is interesting to determine the conditions for it to be interpreted as a Levi-Civita connection of some metric (not necessarily the existing metric on the manifold). These conditions are clearly stated with Christoffel's symbols of first kind, namely 
\be
\pd_\m\bigg(\left\{\d;\b \l\right\}+\left\{\b;\l \d\right\}\bigg)=\pd_\l\bigg(\left\{\d;\b \m \right\}+\left\{\b;\d \m\right\}\bigg)
\ee
which expresses the obvious fact that
\be
\pd_\m\pd_\l~ g_{\d\b}=\pd_\l\pd_\m ~g_{\b\d}
\ee
In order to determine the generaing metric in such cases as it exists, (that is, when the integrability condition is fulfilled), there is the linear ODE system
\be
\pd_\l g_{\d\b}=g_{\a\d} \Gamma^\a_{\b\l}+ g_{\a\b} \Gamma^\a_{\l\d}
\ee
The integrability conditions for such a system are precisely as above, namely

\be
\pd_\m\bigg(g_{\d\a}\Gamma^\a_{\b \l}+g_{\b\a}\Gamma^\a_{\l \d}\bigg)=\pd_\l\bigg(g_{\d\a}\Gamma^\a_{\b \m} +g_{\a\b}\Gamma^\a_{\d \m}\bigg)
\ee
This boils down to
\bea
&&g_{\d\s}\left(\Gamma^\a_{\b\l}\Gamma^\s_{\a\m}-\Gamma^\s_{\l\a}\Gamma^\a_{\b\m}\right)+
g_{\b\s}\left(\Gamma^\a_{\l\d}\Gamma^\s_{\m\a}-\Gamma^\s_{\l\a}\Gamma^\a_{\d\m}\right)
=\left(g_{\d\a}~+g_{\a\b}\right) \left(\pd_\l \Gamma^\a_{\b\m}-~\pd_\m \Gamma^\a_{\l\d}\right)\nonumber
\eea
This can be thought of as a set of {\em algebraic} equations for the metric, given the connection and its derivatives.
At the perturbative level, assuming
\bea
&&g_{\a\b}\equiv \eta_{\a\b}+\kappa h_{\a\b}\nonumber\\
&&\g_{\a\b\g}=O(\kappa)
\eea
The integrability condition reads
\be
\pd_\m\left(\g_{\d\b\l}+\g_{\b\d\l}\right)=\pd_\l\left(\g_{\d\b\m}+\g_{\b\m\d}\right)
\ee
This can be written in a suggestive way as
\be
\pd_\m\g_{\d\b\l}-\pd_\l \g_{\d\b\m}=\pd_\l \g_{\b\d\m}-\pd_\m \g_{\b\d\l}
\ee
or introducing the one-forms
\be
\chi_{\a\b}\equiv \g_{\a\b\l} dx^\l
\ee
this is equivalent to  a certain one-form to be closed, that is,
\be
d~\chi_{(\a\b)}=0
\ee
This implies that
\be
\g_{(\a\b)\l}=\pd_\l \phi_{\a\b}
\ee
Once this condition is fulfilled, the solution is given by the solution of the first order linear differential equation
\be
\pd_\l h_{\d\b}=\g_{\d\b\l}+\g_{\b\d\l}
\ee

\section{Weyl Gravity}

The same property of conformal invariance of the action \eqref{riemannaction} is shared by the Weyl action
\be
S\equiv \int\sqrt{-g}~W_{\m\n\r\s}W^{\m\n\r\s}d^4 x~\label{weylaction}
\ee
where the Weyl tensor is defined as
\bea
&&W^\a_{~\b\g\d}=R[\Gamma]^\a_{~\b\g\d}-{1\over n-2}\left(\d^\a_\g R^+[\Gamma]_{\b\d}-\d^\a_\d R^+[\Gamma]_{\b\g}-\right.\nonumber\\
&&\left.g_{\b\g} R^+[\Gamma]^\a_\d+g_{\b\d} R^+[\Gamma]^\a_\g\right)+{1\over (n-1)(n-2)} R^+[\Gamma]\left(\d^\a_\g g_{\d\b}-\d^\a_\d g_{\g\b}\right)
\eea
With this definition,
\be
W^\l_{~\b\l\d}=0
\ee
but
\be
W^\l_{~\l\a\b}={\cal R}_{\a\b}
\ee
\be
g^{\b\g}W_{\a\b\g\d}=R^+_{\a\d}+R^-_{\a\d}\equiv \hat{R}_{\a\d}
\ee
Besides, due to \eqref{traces}
\be
\hat{R}\equiv g^{\a\d}\hat{R}_{\a\d}=0
\ee

It can be easily shown that it is not possible to modify Weyl's tensor in such a way that it is still antisymmetric in the last two indices and all traces vanish. We shall then refrain from doing any modification on Weyl's tensor.
\par
We would like to insist that at this point (that is, with a connection that is a dynamical variable) there is no real motivation for this particular definition. It is only when the connection is fixed to be the Levi-Civita one that Weyl's tensor acquires its special meaning.
\par

In spite of the fact that Weyl's action does not seem privileged from this point of view, there is a grander viewpoint from which it is. Namely, Cartan's canonical conformal connection \cite{Kobayashi} is a one-form with values in the conformal algebra, $so(2,4)$, in such a way that when two frames are related by a conformal transformation, the connection undergoes a gauge transformation \cite{Attard}. It is a remarkable fact that this connection is closely related to the Penrose's twistor connection \cite{Penrose}.
\par
In this view, spacetime is related  to the coset $SO(2,4)/SO_0(2,4)$ where $SO_0(2,4)$ is the group of scale transformations (that is, the conformal group without the four special conformal transformations).
In \cite{Korzynski} it was shown that Bach's tensor \cite{Bach} is the source of the Yang-Mills' equations for Cartan's connection. Given the well-known fact that  Bach's tensor is the equation of motion for Weyl gravity -in the second order formalism- this clearly yields a new insight. This viewpoint allowed \cite{Wheeler} to show that the solution of the first order Weyl EM were given by conformal classes of solutions of the ordinary Einstein equations. 
 \par
 This is our main motivation for  concentrate on the Weyl action, trying to understand the possible new features that a first order treatment may uncover.
\par
Finally, let us remark {\em en passant}, that the index theorem asserting that the integral of the pfaffian of the curvature yields Euler's characteristic, a topological invariant, refers to the riemannian curvature only. To be specific, there must exist some admissible metric such that the connection is the associated Levi-Civita connection. The conditions for that to be true are worked out in the appendix. There are then in general three independent quadratic diffeomorphism invariants out of the metric and the non-riemannian curvature.
\par
Let us now point out a question of notation to hopefully avoid confusion. When considering the Riemann tensor of the Levi-Civita connection we shall simply write $R^\m_{~\n\r\s}$. When it is computed from a dynamical variable connection, $R^{(\Gamma)\m}_{~\n\r\s}$ and in the important particular case of Weyl's connection (to be introduced in a moment), $R^{(W)\m}_{~\n\r\s}$.

\par
Let us also mention that this very theory (or rather its second order version) has been proposed as experimentally viable in \cite{Kazanas}, although this statement has been challenged by \cite{Flanagan} in the particular case of conformal matter.

\section {Weyl's connection is not enough.}

There is a particular case which is very interesting, namely, when the non-metricity 
\be
Q_{abc}\equiv \nabla_a \eta_{bc}
\ee
({\em confer} the appendix) is proportional to the metric tensor itself
\be
Q_{abc}\equiv -2 W_a ~\eta_{bc}
\ee
where $W_a$ is a gauge field, the Weyl vector field. 
\par
The fact is that when a Weyl transformation is made on the metric
\be
g_{\a\b}\rightarrow \tilde{g}_{\a\b}\equiv \Omega^2 g_{\a\b}
\ee
Weyl's vector undergoes a gauge transformation
\be
\nabla_\m \tilde{g}_{\a\b}=2\left(-W_\m+{\pd_\m\Omega\over\Omega}\right)\tilde{g}_{\a\b}
\ee
In this case it is possible to modify the Levi-Civita connection in such a way that the covariant derivative of the metric with respect to the modified connection still vanishes. Its value is

\begin{align}
\Gamma^{(W)}{}^\mu_{\nu\rho}=\Gamma^{\mu}_{\nu\rho}-\delta^{\mu}_{\nu}W_{\rho}-\delta^{\mu}_{\rho} W_\n +g_{\nu\rho}W^{\mu}
\end{align}
Then when the only non metricity is due to Weyl's vector field,
\be
\nabla_\m g_{\a\b}=\nabla^{(W)}_\m g_{\a\b}
\ee
where $\nabla^{(W)}_\m $ is the covariant derivative corresponding to the Weyl connection, $\Gamma^{(W)}{}^\mu_{\nu\rho}$. 

This fact  allows us to define a conformal (as well as diffeomorphism) covariant derivative of an arbitrary tensor field $T$ through
\begin{align}
D_{\mu}T=\nabla_{\mu}^{(W)}T+\lambda W_{\mu}T
\end{align}
where $\lambda$ is the conformal weight of the tensor T. This is defined in such a way that under a Weyl transformation
\be
T\rightarrow \Omega^{-\l}~T
\ee
  For example, $\l=-2$ for the covariant metric $g_{\a\b}$. Then $D_{\mu}$ is a metric-compatible connection
\be
D_\m g_{\a\b}=0
\ee
The Riemann-Christoffel tensor\footnote{The following formulas correct some unfortunate misprints in \cite{AlvarezHV}
} associated to Weyl's connection reads
\bea
&&R^{(W)}_{\m\n\r\s}=R_{\m\n\r\s}+g_{\m\n}\left(\nabla_\r W_\s-\nabla_\s W_\r\right)+g_{\m\s}\left(\nabla_\r W_\n-W_\r W_\n\right)+\nonumber\\
&&+g_{\m\r}\left(\nabla_\s W_\n-W_\s W_\n\right)+g_{\n\r}\left(\nabla_\s W_\m-W_\s W_\m\right)-g_{\n\s}\left(\nabla_\r W_\m-W_\m W_\r\right)+\nonumber\\
&&+W_\l W^\l\left(g_{\m\s}g_{\n\r}-g_{\m\r} g_{\n\s}\right)
\eea
It has already been pointed out the unfortunate fact that this tensor does not have the property that
\be
R^{(W)}_{\m\n\r\s}=R^{(W)}_{\r\s\m\n}
\ee
in fact, it is not even symmetric in the first two indices. As we already advertised in the Introduction, there are then two possible independent contractions of it. 
The Ricci tensor corresponds to the contraction of the first and the third indices and reads
\bea
&&R_{\n\s}^{+(W)}=R_{\n\s}+\nabla_\n W_\s -(n-1)\nabla_\s W_\n- g_{\n\s}\left(\nabla_\l W^\l+(n-2)W_\l W^\l\right)\nonumber
\eea
The other two-index tensor corresponds to the contraction of the second and third indices and read
\bea
&&R^{-(W)}_{\m\s}=-R_{\m\s}+\nabla_\m W_\s+(n-3)\nabla_\s W_\m-3 W_\s W_\m + g_{\m\s}\left((n-2)W_\l W^\l +\nabla_\l W^\l\right)\nonumber
\eea
Finally, the corresponding scalars read
\be
R^{+(W)}=R-2(n-1)\nabla_\l W^\l-(n-2)(n-1) W_\l W^\l=-R^{-(W)}
\ee

Let us denote by $F_{\a\b}$ the gauge invariant field strength of the abelian gauge field $W_\l$
\be
F_{\a\b}\equiv \nabla_\a W_\b-\nabla_\b W_\a=\pd_\a W_\b-\pd_\b W_\a
\ee
It is easy to check that the contraction defined in \eqref{ricciblando}
\be
{\cal R}_{\r\s}=(n-2) F_{\r\s}
\ee
 Weyl's tensor corresponding to Weyl's connection reads
\bea
&&W^{(W)}_{\m\n\r\s}=W_{\m\n\r\s}+ g_{\m\n}F_{\r\s}+{1\over n-2}\left(g_{\m\r}F_{\s\n}-g_{\m\s}F_{\r\n}-g_{\n\r} F_{\s\m}+g_{\n\s}F_{\r\m}\right)\nonumber
\eea
which is indeed gauge invariant in a beautiful way.
\par
Unfortunately this also means that the lagrangian corresponding to Weyl gravity where the connection only depends on the Weyl gauge field can be written as
\be
\sqrt{|g|}~W^{(W)}_{\a\b\g\d} W_{(W)}^{\a\b\g\d}=\sqrt{|g|}\bigg\{W_{\m\n\r\s}W^{\m\n\r\s}+{n^2-2n+8\over n-2}F_{\m\n}~F^{\m\n}\bigg\}
\ee
This means that the whole dynamics of Weyl's connection lies in the abelian gauge term and the first order formalism for the gravitational field has got exactly the same problems with unitarity as has the usual second  order one.

\section{Conclusions}
It has been known for some time that the complete on shell equivalence of first order (Palatini) and second order formalisms for the Einstein-Hilbert action does not hold for quadratic actions. This is an inducement to investigate those theories anew, because by considering the connection and the metric as independent dynamical variables, there are no quartic propagators for any variable, at least naively. These theories then hold the promise of defining a unitary and renormalizable theory of gravity.
Given the quite stringent observational bounds on the presence of torsion in spacetime, we have assumed throughout the paper that the torsion is zero, although this hypothesis could easily be removed.
\par

We have clarified the general setup of the theory, and the degrees of freedom involved.
\par
In this paper we have endeavored to work out the dynamics of Weyl's connection, which has a very natural and beautiful geometrical interpretation. Unfortunately, for this  particular case  in which the non-metricity is proportional to the metric itself  the propagator for the graviton is still quartic in derivatives, which means that in momentum space behaves as 
$\tfrac{1}{p^4}$ for large momentum. 
\par
Our conclusion is that those theories are still non-unitary even in first order formalism.
\par
This means that the only hope for getting a renormalizable and unitary theory involving gravitons (alas, also other fields) lies in the general quartic lagrangian, which unfortunately is quite complicated because it depends on many parameters (one for every non-equivalent way of contracting the Riemann tensor with itself). 
\par
We hope to be able to report on this sopic soon.
\section*{Acknowledgments}
We are grateful to the LBNL and UC Berkeley  for hospitality in the last stages of this work.
We acknowledge many enjoyable and informative discussions with JLF Barb\'on, CP Mart\'in, Gonzalo Olmo and Jes\'us Anero as well as useful e-mail exchange with John F. Donoghue and Jose Senovilla. This project has received funding from the European Union's Horizon 2020 research and innovation programme under the Marie Sklodowska-Curie grants agreement No 674896 and No 690575. We also have been partially supported by FPA2012-31880 (Spain), COST actions MP1405 (Quantum Structure of Spacetime) and  COST MP1210 (The string theory Universe).  The authors acknowledge the support of the Spanish MINECO {\em Centro de Excelencia Severo Ochoa} Programme under grant  SEV-2012-0249. We are grateful to Stanley Deser and to  Mike Duff for comments on the manuscript.

\newpage

\appendix
\section{Proof of Lanczos'identity}
Let us now present for completeness Lovelock's proof of the Lanczos'identity \cite{Lanczos}, which is much simpler that the original proof.
\par
 Consider a tensor
\be
K_{\a_1\ldots\a_m}^{\b_1\ldots\b_m}
\ee
such that it is fully traceless and separately antisymmetric in both covariant and contravariant indices. This a generalization of the well-known properties of the Weyl tensor (when the connection is Levi-Civita). Now, if the dimension of the space
\be
n\leq 2~m-1
\ee
then
\be
K_{\a_1\ldots\a_m}^{\b_1\ldots\b_m}\equiv 0
\ee
Define the auxiliary tensor
\be
T_{\a_1\ldots\a_m}^{\b_1\ldots\b_m}\equiv \d _{\a_1\ldots\a_m\g_1\ldots\g_m}^{\b_1\ldots\b_m\d_1\ldots\d_m}~K_{\d_1\ldots\d_m}^{\g_1\ldots\g_m}
\ee
It is not difficult to check that (because of the conditions assumed on K)
\be
T_{\a_1\ldots\a_m}^{\b_1\ldots\b_m}\equiv \d^{\b_1\ldots\b_m}_{\a_1\ldots\a_m}\d_{\g_1\ldots\g_m}^{\d_1\ldots\d_m}~K_{\d_1\ldots\d_m}^{\g_1\ldots\g_m}=\left(m!\right)^2~K_{\a_1\ldots\a_m}^{\b_1\ldots\b_m}
\ee

But Kronecker's tensor vanishes unless $n\geq 2m$, which is what we wanted to prove.
\par
When the dimension is even, $n=2m$,  this theorem then implies a second one, namely that
\be
K_{[\a_1\ldots\a_m}^{[\b_1\ldots\b_m}\d^{\b_{m+1}]}_{\a_{m+1}]}=0
\ee
The reason is that this particular tensor fulfills all the hypothesis of the preceding theorem. Let us check this explicitly for the Weyl tensor, which corresponds to $m=2$. This construct reads
\bea
&&{1\over 9}\bigg\{W_{~~\m\n}^{\a\b}\d_\l^\g+W_{~~\m\n}^{\b\g}\d_\l^\a+W_{~~\m\n}^{\g\a}\d_\l^\b+
W_{~~\n\l}^{\a\b}\d_\m^\g+W_{~~\n\l}^{\b\g}\d_\m^\a+W_{~~\n\l}^{\g\a}\d_\m^\b+\nonumber\\
&&+
W_{~~\l\m}^{\a\b}\d_\n^\g+W_{~~\l\m}^{\b\g}\d_\n^\a+W_{~~\l\m}^{\g\a}\d_\n^\b\bigg\}=0
\eea
It is easily checked that all traces indeed vanish. If we now multiply the ensuing identity
by
\be
W^{\m\n}_{~~\a\b}
\ee
we obtain the desired result
\be
W_{~~\m\n}^{\a\b}~W^{\m\n}_{~~\a\b}~\d_\r^\s=4~W^{~~\m\n}_{\a\r}~W_{\m\n}^{~~\a\s}
\ee
In conclusion, we are not aware of any  argument as to why Lanczos' identity should remain true for non Levi-Civita connections.


\end{document}